\documentclass[english,prd,showpacs,showkeys,preprintnumbers]{revtex4}
\usepackage{amsmath}
\usepackage{graphicx}
\usepackage{amssymb}
\usepackage{esint}

\makeatletter

\usepackage{xcolor}

\PassOptionsToPackage{normalem}{ulem}\usepackage{ulem}\usepackage[unicode=true, pdfusetitle,
 bookmarks=true,bookmarksnumbered=false,bookmarksopen=false,
 breaklinks=false,pdfborder={0 0 1},backref=false,colorlinks=false]{hyperref}

\makeatother

\usepackage{babel}

\begin{document}

\title{DDVCS on nucleons and nuclei}

\author{B.Z.~Kopeliovich}

\email{Boris.Kopeliovich@usm.cl}

\author{Ivan~Schmidt}

\email{Ivan.Schmidt@usm.cl}

\author{M.~Siddikov}

\email{Marat.Siddikov@usm.cl}

\affiliation{Departamento de F\'{\i}sica, Centro de Estudios Subat\'omicos, y Centro Cient\'ifico - Tecnol\'ogico de Valpara\'iso, Universidad T\'ecnica Federico Santa Mar\'{\i}a, Casilla 110-V, Valpara\'iso, Chile}

\preprint{USM-TH-264}
\begin{abstract}
In this paper we evaluate the double deeply virtual Compton scattering
on nucleons and nuclei in the framework of the color dipole model.
Both the effects of quark and the gluon shadowing are taken into account. 
\end{abstract}

\keywords{Double deeply virtual Compton scattering, color dipole model}

\pacs{13.60.Fz, 24.85.+p, 12.40.-y}

\maketitle

\section{Introduction}

Compton scattering, $\gamma^{*}+p\to\gamma+p$, with initial photons
real or virtual, has been intensively investigated theoretically and
experimentally \cite{Mueller:1998fv,Ji:1996nm,Ji:1998pc,Radyushkin:1996nd,Radyushkin:1997ki,Radyushkin:2000uy,Ji:1998xh,Collins:1998be,Collins:1996fb,Brodsky:1994kf,Goeke:2001tz,Diehl:2000xz,Belitsky:2001ns,Diehl:2003ny,Belitsky:2005qn,Goncalves:2010ci,Boffi:2007yc}.
In the case of deeply-virtual Compton scattering (DVCS), where the
initial photon is highly-virtual, the QCD factorization has been proven
\cite{Radyushkin:1997ki,Collins:1998be,Ji:1998xh} and the amplitude
can be expressed in terms of the generalized parton distributions
(GPD) \cite{Mueller:1998fv,Ji:1996nm,Ji:1998pc,Radyushkin:1996nd,Radyushkin:1997ki,Radyushkin:2000uy,Ji:1998xh,Collins:1998be,Collins:1996fb,Brodsky:1994kf,Goeke:2001tz,Diehl:2000xz,Belitsky:2001ns,Diehl:2003ny,Belitsky:2005qn}
convoluted with some hard coefficient function. However, it is not
possible to make a deconvolution and unambiguously extract the GPDs
of the target from the experimental DVCS data. The standard procedure
in this case is to construct a plausible model with some free parameters,
fix the free parameters from the experimental data and after that
make conclusions about the GPDs of the target.

The situation is different in case of double deeply virtual Compton
scattering (DDVCS), $\gamma^{*}+p\to\gamma^{*}+p\to\bar{l}l+p$~\cite{Guidal:2002kt,Belitsky:2002tf,Belitsky:2003fj}.
As it has been shown in~\cite{Guidal:2002kt}, the DDVCS amplitude
allows such a deconvolution. Unfortunately, the corresponding cross-section
of the process is suppressed by $\alpha_{em}/3\pi$ compared to ordinary
DVCS and falls into a picobarn-level range, so it cannot be accessed
by the existing accelerating facilities. However, at future accelerating
facilities \cite{Klein:2008zza,Zimmermann:2008zzd} with higher luminosities,
this cross-section can be measured. A special case, when the initial
photon is real, has been studied in~\cite{Berger:2001xd,Pire:2008ea},
and in the color dipole framework in \cite{Machado:2008zv}. This
process may be viewed as time-inverted DVCS with negative virtuality
$-M_{\bar{l}l}^{2}$, so the color dipole model is applicable for
high lepton masses, $M_{\bar{l}l}^{2}\gg M_{N}^{2}.$ A high luminosity
flux of real photons may be created by ultraperipheral scattering
of protons on nuclei; such collisions at the LHC will be an ideal
tool for the study of this process. However, it is not possible to
access the whole DDVCS amplitude in ultraperipheral hadronic collisions,
since the virtuality of the initial photon, controlled by the formfactors
of the colliding hadrons, cannot be very high. Thus, the only way
to study the DDVCS is the electroproduction of lepton pairs, $e+p\to e+p+\bar{l}l$.

Recently DVCS on proton and nuclear targets has been studied within
the color dipole approach in~\cite{Kopeliovich:2008ct,Kopeliovich:2009cx,Machado:2008tp,Machado:2009cd,Kopeliovich:2010sa}.
In this paper we extend that study to the DDVCS case. We focus on
the high-energy kinematics range, since the future accelerating facilities
are expected to operate at these energies~\cite{Klein:2008zza,Zimmermann:2008zzd},
where the effects of coherence are important. The general framework
for evaluating the shadowing corrections is the Gribov-Glauber approach~\cite{Gribov:1968jf,Glauber:1955qq}.
In particular, we take into account gluon shadowing corrections, which
correspond to the triple-Pomeron diffraction in the Gribov inelastic
shadowing corrections. These corrections onset at $x_{B}\lesssim10^{-2}$,
and give a sizeable contribution to nuclear shadowing at $x_{B}\sim10^{-5}$.

The paper is organized as follows. In Sections~\ref{sec:DipoleModel}
we review the general formalism of the color dipole approach. In Section~\ref{sec:frozen}
we discuss the frequently used frozen approximation, which is valid
for asymptotically large energies. In Section~\ref{sec:sha-coh}
we discuss the method which will be used for calculations of nuclear
shadowing effects and demonstrate that for asymptotically large energies
it reproduces the results from~Section~\ref{sec:frozen}. In Section~\ref{sec:GluonShadowing}
we discuss the gluon shadowing and its effect on the DDVCS observables.
In Section~\ref{sec:WFfromIVM} the wavefunction of a real photon
is evaluated in the instanton vacuum model. In Section~\ref{sec:Results}
we present the results of numerical evaluation and draw conclusions.

\section{Color dipole model}

\label{sec:DipoleModel} The electroproduction cross-section for unpolarized
lepton pairs in the DDVCS process has a form

\begin{eqnarray}
\frac{d\sigma^{\bar{l}lp}}{dt\, dM_{\bar{l}l}^{2}} & = & \frac{d\sigma_{el}^{\gamma p}}{dt}\frac{\alpha_{em}}{3\pi M_{\bar{l}l}^{2}}\left(1-\frac{4m_{l}^{2}}{M_{\bar{l}l}^{2}}\right)^{3/2},\label{eq:DDVCS-corss-section}\end{eqnarray}
where $d\sigma_{el}^{\gamma p}/dt$ is the cross-section of the process
$\gamma^{*}p\to\gamma^{'*}p$, $M_{\bar{l}l}^{2}$ is the invariant
mass of the produced lepton pair, and $t$ is the square of the momentum
transferred to a target. In what follows, we will concentrate on evaluation
of the cross-section $d\sigma_{el}^{\gamma p}/dt$. For the unpolarized
Compton scattering discussed in this paper we obtain, \begin{eqnarray}
\frac{d\sigma_{el}^{\gamma p}}{dt} & = & \frac{1}{16\pi}\sum_{ij}\left|\mathcal{A}_{\mu\nu}^{(ij)}\right|^{2},\label{eq:DVCS-cross-section-2}\end{eqnarray}

where the amplitude $\mathcal{A}_{\mu\nu}^{(ij)}$ is evaluated in
the color dipole model. In this model the dominant contribution to
the Compton amplitude comes from gluonic exchanges. Then the general
expression for the Compton amplitude on a nucleon has the form,

\begin{equation}
\mathcal{A}_{\mu\nu}^{(ij)}\left(s,\Delta,Q^{2},M_{\bar{l}l}^{2}\right)=e_{\mu}^{(i)}e_{\nu}^{(j)}\int\limits _{0}^{1}d\beta_{1}d\beta_{2}d^{2}r_{1}d^{2}r_{2}\bar{\Psi}_{f}^{(i)}\left(\beta_{2},\vec{r}_{2}\right)\mathcal{A}^{d}\left(\beta_{1},\vec{r}_{1};\beta_{2},\vec{r}_{2};\Delta\right)\Psi_{in}^{(j)}\left(\beta_{1},\vec{r}_{1}\right),\label{eq:Convolution:Full}\end{equation}
 where $e_{\mu}^{(i)}$ is the photon polarization vector; $\beta_{1,2}$
are the light-cone fractional momenta of the quark and antiquark,
$\vec{r}_{1,2}$ are the transverse distances in the final and initial
dipoles respectively; $\Delta$ is the momentum transfer in the Compton
scattering\textbf{,} $\mathcal{A}^{d}(...)$ is the scattering amplitude
of the dipole on the target (proton or nucleus), and $\Psi_{in(f)}^{(i)}\left(\beta,\vec{r}\right)$
are the light-cone distribution functions of the initial and final
photons in the polarization state $i$ . When the virtuality of the
corresponding photon is large, we may use the well-known pQED expressions~\cite{Kogut:1969xa,Bjorken:1970ah},\[
\bar{\Psi}_{f}^{(i)}\Psi_{in}^{(j)}\sim\left(\left(\alpha^{2}+(1-\alpha)^{2}\right)\epsilon_{1}\epsilon_{2}K_{1}(\epsilon_{1}r)K_{2}(\epsilon_{2}r)+m_{q}^{2}K_{0}(\epsilon_{1}r)K_{0}(\epsilon_{2}r)\right),\]
 where $\epsilon_{i}=\sqrt{Q_{i}^{2}\alpha(1-\alpha)+m_{q}^{2}}$,
and make an analytical continuation of the space-like wave functions
to the time-like photons by a simple extrapolation, $Q^{2}\to-M_{\bar{l}l}^{2}$~\cite{Motyka:2008ac},
and use \begin{align}
K_{0}(ix) & =-\frac{i\pi}{2}\left(J_{0}(x)-iY_{0}(x)\right),\label{eq:McDonald_Im_1}\\
K_{1}(ix) & =-\frac{\pi}{2}\left(J_{1}(x)-iY_{1}(x)\right),\label{eq:McDOnald_Im_2}\end{align}
 where $x=r\sqrt{M_{\bar{l}l}^{2}\alpha(1-\alpha)-m_{q}^{2}}$ for
the kinematics $M_{\bar{l}l}^{2}\gg m_{q}^{2}/(\alpha(1-\alpha))$.
However when the virtualities are small, we have to resort to some
model (see Section~\ref{sec:WFfromIVM} for more details).

At high energies in the small angle approximation, $\Delta/\sqrt{s}\ll1$,
the quark separation and fractional momenta $\beta$ are preserved,
so \begin{eqnarray}
\mathcal{A}^{d}\left(\beta_{1},\vec{r}_{1};\beta_{2},\vec{r}_{2};Q^{2},\Delta\right) & \approx & \delta\left(\beta_{1}-\beta_{2}\right)\delta\left(\vec{r}_{1}-\vec{r}_{2}\right)\int d^{2}b'\, e^{i\vec{\Delta}\vec{b}'}\Im m\, f_{\bar{q}q}^{N}(\vec{r}_{1},\vec{b}',\beta_{1})\label{eq:DVCSIm-BK}\\
\Im m\, f_{\bar{q}q}^{N}(\vec{r},\vec{b}',\beta) & = & \frac{1}{12\pi}\int\frac{d^{2}k\, d^{2}\Delta}{\left(k+\frac{\Delta}{2}\right)^{2}\left(k-\frac{\Delta}{2}\right)^{2}}\alpha_{s}\mathcal{F}\left(x,\vec{k},\vec{\Delta}\right)\, e^{i\vec{b}'\cdot\vec{\Delta}}\nonumber \\
 & \times & \left(e^{-i\beta\vec{r}\cdot\left(\vec{k}-\frac{\vec{\Delta}}{2}\right)}-e^{i(1-\beta)\vec{r}\cdot\left(\vec{k}-\frac{\vec{\Delta}}{2}\right)}\right)\left(e^{i\beta\vec{r}\cdot\left(\vec{k}+\frac{\vec{\Delta}}{2}\right)}-e^{-i(1-\beta)\vec{r}\cdot\left(\vec{k}+\frac{\vec{\Delta}}{2}\right)}\right),\label{eq:ImFN_GPD}\end{eqnarray}
 where \begin{eqnarray}
\frac{\mathcal{F}\left(x,\vec{k},\vec{\Delta}\right)}{k^{2}} & \equiv & H_{g}\left(x,\vec{k},\vec{\Delta}\right)=\frac{1}{2}\int d^{2}r\, e^{ik\cdot r}\int\frac{dz^{-}}{2\pi}e^{ix\bar{P}^{+}z^{-}}\times\nonumber \\
 & \times & \left\langle P'\left|G_{+\alpha}\left(-\frac{z}{2},\,-\frac{\vec{r}}{2}\right)\gamma_{+}\mathcal{L}\left(-\frac{z}{2}-\frac{\vec{r}}{2},\frac{z}{2}+\frac{\vec{r}}{2}\right)G^{+\alpha}\left(\frac{z}{2},\,\frac{\vec{r}}{2}\right)\right|P\right\rangle \label{eq:UninteggratedGPD-definition}\end{eqnarray}
 is the gluon GPD of the target, $P'=P+\Delta,$ $\bar{P}=(P+P')/2$,
$G_{\mu\nu}(x)$ is the gluon loop operator, $\mathcal{L}_{\infty}\left(x,y\right)$
is the Wilson factor required by gauge covariance. For this GPD we
use a gaussian parameterization~\cite{Kopeliovich:2007fv,Kopeliovich:2008nx,Kopeliovich:2008da},

\begin{equation}
\mathcal{F}\left(x,\vec{k},\vec{\Delta}\right)=\frac{3\sigma_{0}(x)}{16\pi^{2}\alpha_{s}}\left(k+\frac{\Delta}{2}\right)^{2}\left(k-\frac{\Delta}{2}\right)^{2}R_{0}^{2}(x)\exp\left(-\frac{R_{0}^{2}(x)}{4}\left(\vec{k}^{2}+\frac{\vec{\Delta}^{2}}{4}\right)\right)\exp\left(-\frac{1}{2}B(x)\Delta^{2}\right),\label{eq:GPDParametrization}\end{equation}
 where the phenomenological functions $\sigma_{0}(x),\, R_{0}^{2}(x),\, B(x)$
are fitted to DIS, real photoproduction and $\pi p$ scattering data.
We discuss them in more detail in Section~\ref{sec:Results}. The
parameterization~(\ref{eq:GPDParametrization}) does not depend on
the longitudinal momentum transfer and decreases exponentially as
a function of $\Delta^{2}$. Since the parameterization~(\ref{eq:GPDParametrization})
is an effective one and is valid only in the small-$x$ region, we
do not assume that it satisfies general requirements, such as positivity~\cite{Pobylitsa:2002gw}
and polynomiality~\cite{Ji:1998pc} constraints.

The prefactor $\left(k+\frac{\Delta}{2}\right)^{2}\left(k-\frac{\Delta}{2}\right)^{2}$
in~(\ref{eq:GPDParametrization}) guarantees convergence of the integrals
in the parameterization~(\ref{eq:DVCSIm-BK}). In the forward limit,
$\Delta\to0$, the amplitude (\ref{eq:DVCSIm-BK}) reduces to the
saturated parameterization of the dipole amplitude proposed by Golec-Biernat
and W\"usthoff (GBW)~\cite{GolecBiernat:1998js}, \begin{eqnarray}
\sigma_{d}(\beta,r) & = & 2\int d^{2}b'\,\Im m\, f_{\bar{q}q}^{N}(\vec{r},\vec{b}',\beta)=\frac{1}{6\pi}\int\frac{d^{2}k}{k^{4}}\alpha_{s}\left(k^{2}\right)\mathcal{F}\left(x,\vec{k},\vec{0}\right)=\frac{\sigma_{0}(x)}{2}\left(1-\exp\left(-\frac{r^{2}}{R_{0}(x)}\right)\right)\label{eq:Sigma_QQ_GPD}\end{eqnarray}
 Generally, the amplitude $f_{\bar{q}q}^{N}(...)$ involves nonperturbative
physics, but its asymptotic behavior for small~$r$ is controlled
by pQCD \cite{Kopeliovich:1981}:\[
f_{\bar{q}q}^{N}(\vec{r},\vec{\Delta},\beta)_{r\to0}\propto r^{2},\]
 up to slowly varying corrections $\sim\ln(r)$.

The calculation of the differential cross section also involves the
real part of scattering amplitude, whose relation to the imaginary
part is quite straightforward. According to \cite{Bronzan:1974jh},
if the limit $\lim\limits _{s\to\infty}\left(\frac{\mathcal{I}m\,{f}}{s^{\alpha}}\right)$
exists and is finite, then the real and imaginary parts of the forward
amplitude are related as\begin{equation}
\mathcal{R}e\,{f(\Delta=0)}=s^{\alpha}\tan\left[\frac{\pi}{2}\left(\alpha-1+\frac{\partial}{\partial\ln s}\right)\right]\frac{\Im m\,{f(\Delta=0)}}{s^{\alpha}}.\label{eq:BronzanFul}\end{equation}
 In the model under consideration the imaginary part of the forward
dipole amplitude indeed has a power dependence on energy, $\mathcal{I}m\, f(\Delta=0;\, s)\sim s^{\alpha}$,
so (\ref{eq:BronzanFul}) simplifies to \begin{eqnarray}
\frac{\mathcal{R}e\,\mathcal{A}}{\Im m\,\mathcal{A}} & =\tan\left(\frac{\pi}{2}(\alpha-1)\right)\equiv\epsilon.\end{eqnarray}

This fixes the phase of the forward Compton amplitude, which we retain
for nonzero momentum transfers, assuming similar dependences for the
real and imaginary parts. Finally we arrive at, \begin{equation}
\mathcal{A}_{\mu\nu}^{(ij)}=(\epsilon+i)e_{\mu}^{(i)}(q')e_{\nu}^{(j)}(q)\int d^{2}r\int\limits _{0}^{1}d\beta\,\bar{\Psi}_{f}^{(i)}(\beta,r)\Psi_{in}^{(j)}(\beta,r)\,\Im m\, f_{\bar{q}q}^{N}(\vec{r},\vec{\Delta},\beta,s),\label{eq:DVCS-Im-conv}\end{equation}

\section{Nuclear shadowing in the frozen limit}

\label{sec:frozen}

Nuclear shadowing signals the closeness of the unitarity limit. Hard
reactions possess this feature only if they have a contribution from
soft interactions. In DIS and DVCS the soft contribution arises from
the so called aligned jet configurations \cite{bjorken}, corresponding
to $\bar{q}q$ fluctuations very asymmetric in sharing the photon
momentum, $\beta\ll1$. Such virtual photon fluctuations, having large
transverse separation, are the source of shadowing \cite{k-povh}
.

Calculation of nuclear shadowing simplifies considerably in the case
of long coherence length \cite{krt2}, i.e. long lifetime of the photon
fluctuations, when it considerably exceeds the nuclear size. In this
case Lorentz time dilation "freezes" the
transverse size of the fluctuation during propagation though the nucleus.
Then the Compton amplitude of coherent scattering, which leaves the
nucleus intact, has the same form as Eq.~(\ref{eq:DVCS-Im-conv})
with a replacement of the nucleon Compton amplitude by the nuclear
one, \begin{equation}
\Im mf_{\bar{q}q}^{N}(r,\beta,\Delta)\Rightarrow{\cal \Im}mf_{\bar{q}q}^{A}(r,\beta,\Delta)=\int d^{2}b\, e^{i\vec{\Delta}\cdot\vec{b}}\,\left[1-e^{-{\cal \Im}mf_{\bar{q}q}^{N}(r,\beta,\Delta=0)\, T_{A}(b)}\right],\label{nucl-coh}\end{equation}
 where $b$ is impact parameter of the photon-nucleus collision, $T_{A}(b)=\int_{-\infty}^{\infty}dz\,\rho_{A}(b,z)$
is the nuclear thickness function, given by the integral of nuclear
density along the direction of the collisions. In this expression
we neglect the real part of the amplitude which is particularly small
for a coherent nuclear interaction.

For incoherent Compton scattering, which results in nuclear fragmentation
without particle production (quasielastic scattering), the cross section
has the form \cite{kps1},

\begin{eqnarray}
\frac{d\sigma_{qel}^{\gamma A}}{dt} & = & B_{el}\, e^{B_{el}t}\,\sum_{ij}\int\limits _{0}^{1}d\beta\int d^{2}rd^{2}r'\,\bar{\Psi}_{f}^{(i)}(\beta,r)\,\bar{\Psi}_{f}^{(i)}(\beta,r')\,\Psi_{in}^{(j)}(\beta,r)\,\Psi_{in}^{(j)}(\beta,r')\nonumber \\
 & \times & \exp\left[\frac{1}{2}\Bigl(\sigma_{\bar{q}q}(r)-\sigma_{\bar{q}q}(r')\Bigr)\, T_{A}(b)\right]\,\left\{ \exp\left[\frac{\sigma_{\bar{q}q}(r)\,\sigma_{\bar{q}q}(r')}{16\pi\, B_{el}}\, T_{A}(b)\right]-1\right\} \nonumber \\
 & \approx & \frac{e^{B_{el}t}}{16\pi}\int d^{2}b\, T_{A}\left(b\right)\left|\int\limits _{0}^{1}d\beta\, d^{2}r\,\bar{\Psi}_{f}\left(\beta,r\right)\sigma_{\bar{q}q}\left(r,s\right)\Psi_{in}\left(\beta,r\right)\exp\left[-\frac{1}{2}\sigma_{\bar{q}q}\left(r\right)T_{A}\left(b\right)\right]\right|^{2}.\label{eq:quasielastic}\end{eqnarray}
 Here $B_{el}$ is the $t$-slope of elastic dipole-nucleon amplitude.
In this equation we treated the term quadratic in the dipole cross
section as a small number and expanded the exponential in curly brackets.

\section{Onset of nuclear shadowing}

\label{sec:sha-coh}

The regime of frozen dipole size discussed in the previous section
is valid only at very small $x_{B}$, or at high energies. However,
at medium small $x_{B}$ a dipole can "breath",
i.e. vary its size, during propagation through the nucleus, and one
should rely on a more sophisticated approach.

In this paper we employ the description of the onset of shadowing
developed in~\cite{Kopeliovich:1998gv} and based on the light-cone
Green function technique \cite{kz91}. The propagation of a color
dipole in a nuclear medium is described as motion in an absorptive
potential, \emph{i.e.} \begin{equation}
i\,\frac{\partial G\left(z_{2},r_{2};z_{1},r_{1}\right)}{\partial z_{2}}=-\frac{\Delta_{r_{2}}G\left(z_{2},r_{2};z_{1},r_{1}\right)}{\nu\alpha(1-\alpha)}-k_{min}G\left(z_{2},r_{2};z_{1},r_{1}\right)-\frac{i\rho_{A}\left(z_{2},r_{2}\right)\sigma_{\bar{q}q}\left(r_{2}\right)}{2}\, G\left(z_{2},r_{2};z_{1},r_{1}\right),\label{eq:W-definition}\end{equation}
 where the Green function $G\left(z_{2},r_{2};z_{1},r_{1}\right)$
describes the probability amplitude for the propagation of dipole
state with size $r_{1}$ at the light-cone starting point $z_{1}$
to the dipole state with size $r_{2}$ at the light-cone point $z_{2},$
and \[
k_{min}=\frac{Q^{2}\alpha(1-\alpha)+m_{q}^{2}}{2\nu\alpha(1-\alpha)}.\]
 Then the shadowing correction to the amplitude has the form\begin{eqnarray}
\delta A(s,\vec{\Delta}_{\perp}) & = & \int d^{2}b\, e^{i\vec{\Delta}_{}\cdot\vec{b}_{\perp}}\int\limits _{z_{1}\le z_{2}}dz_{1}dz_{2}\,\rho_{A}(b,z_{1})\,\rho_{A}(b,z_{2})\int\limits _{0}^{1}d\alpha d^{2}r_{1}d^{2}r_{2}\times\nonumber \\
 &  & \bar{\Psi}_{f}\left(\alpha,r_{2}\right)\sigma_{\bar{q}q}\left(r_{2}\right)\, G\left(z_{2},r_{2};z_{1},r_{1}\right)\sigma_{\bar{q}q}\left(r_{1}\right)\Psi_{in}\left(\alpha,r_{1}\right).\label{eq:correction}\end{eqnarray}

Equation~(\ref{eq:W-definition}) is quite complicated and in the
general case may be solved only numerically \cite{nemchik}. However
in some cases an analytic solution is possible. For example, in the
limit of long coherence length, $l_{c}\gg R_{A}$, relevant for high-energy
accelerators like the LHC, one can neglect the {}``kinetic'' term~$\propto\Delta_{r_{2}}G\left(z_{2},r_{2};z_{1},r_{1}\right)$
in ~(\ref{eq:W-definition}), and get the Green function in the "frozen"
approximation \cite{kz91}, \begin{equation}
G\left(z_{2},r_{2};z_{1},r_{1}\right)=\delta^{2}\left(r_{2}-r_{1}\right)\exp\left(-\frac{1}{2}\sigma_{\bar{q}q}\left(r_{1}\right)\int\limits _{z_{1}}^{z_{2}}d\zeta\,\rho_{A}\left(\zeta,b\right)\right)e^{ik_{min}\left(z_{2}-z_{1}\right)}.\label{eq:W-frozen}\end{equation}
 Then the shadowing correction~(\ref{eq:correction}) simplifies
to \begin{eqnarray}
\delta\mathcal{A}(s,\Delta_{\perp}) & = & \int d^{2}b\, e^{i\vec{\Delta}_{\perp}\cdot\vec{b}_{\perp}}\int\limits _{z_{1}\le z_{2}}dz_{1}dz_{2}\,\rho_{A}\left(z_{1},b\right)\rho_{A}\left(z_{2},b\right)\int\limits _{0}^{1}d\alpha\, d^{2}r\,\sigma_{\bar{q}q}^{2}\left(r,b\right)\times\label{eq:DVCS-COH-SHA}\\
 &  & \bar{\Psi}_{f}\left(\alpha,r\right)\exp\left(-\frac{1}{2}\sigma_{\bar{q}q}\left(r\right)\int\limits _{z_{1}}^{z_{2}}d\zeta\rho_{A}\left(\zeta,b\right)\right)\Psi_{in}\left(\alpha,r\right)e^{ik_{min}\left(z_{2}-z_{1}\right)}.\nonumber \end{eqnarray}
 If we neglect the real part of the amplitude and the longitudinal
momentum transfer $k_{min}$ (which is justified for asymptotically
large $s$), and average over polarizations, then taking the integral
over $z_{1,2}$ ''by parts'' in~(\ref{eq:DVCS-COH-SHA}), we get
for the elastic amplitude \begin{eqnarray}
\mathcal{A}(s,\Delta_{\perp}) & = & 2\int d^{2}b\, e^{i\vec{\Delta}_{\perp}\cdot\vec{b}_{\perp}}\int\limits _{0}^{1}d\alpha\, d^{2}r\,\bar{\Psi}_{f}\left(\alpha,r\right)\left[1-\exp\left(-\frac{1}{2}\sigma_{\bar{q}q}\left(r\right)\int\limits _{-\infty}^{+\infty}d\zeta\rho_{A}\left(\zeta,b\right)\right)\right]\Psi_{in}\left(\alpha,r\right).\label{eq:A-frozen}\end{eqnarray}
 Another case where an analytical solution is possible is when the
effective dipole sizes are small and the function $\sigma_{\bar{q}q}(r)$
may be approximated as \begin{equation}
\sigma_{\bar{q}q}(r)\approx Cr^{2}.\label{eq:sigma-small-r}\end{equation}
 This approximation cannot be precise even at high virtualities $Q^{2}$
and $M_{\bar{l}l}^{2}$ in DDVCS, since there are contributions of
the aligned jet configurations mentioned above, which permit large
dipoles even for large virtualities. Moreover, such aligned jet configurations
of the dipole provide the main contribution to nuclear shadowing \cite{k-povh}.
Nevertheless, for the sake of simplicity we use this approximation
in order to \emph{estimate} the magnitude of the shadowing corrections
in the region $x_{B}\in\left(10^{-3},10^{-1}\right)$. The approximation~(\ref{eq:sigma-small-r})
is well justified on heavy nuclei. Namely, nuclear shadowing is independent
of the form of the dipole cross section for large dipole sizes, above
the saturation point, $r^{2}>4/Q_{s}^{2}$, where the typical value
of the saturation momentum is $Q_{s}^{2}\sim1\, GeV^{2}$ for heavy
nuclei. Indeed, in this case the nucleus is "black".
Therefore the shape of the dipole cross section matters only at $r^{2}<r_{s}^{2}=0.16\, fm^{2}$.
This size is sufficiently small for using the $r^{2}$ approximation~(\ref{eq:sigma-small-r}).
Numerically, the approximation~(\ref{eq:sigma-small-r}) was tested
in~\cite{nemchik}--it was found that the discrepancy between the
approximation~(\ref{eq:sigma-small-r}) and the exact numerical solution
of~(\ref{eq:W-definition}), changes the nuclear shadowing for DIS
only within ten percent. We expect that within the same accuracy the
approximation~(\ref{eq:sigma-small-r}) is valid for DVCS.

Then Eq.~(\ref{eq:W-definition}) yields for $W\left(z_{2},r_{2};z_{1},r_{1}\right)$
the well-known evolution operator of harmonic oscillator, although
with complex frequency \begin{eqnarray}
G\left(z_{2},r_{2};z_{1},r_{1}\right) & = & \frac{a}{2\pi i\sin\left(\omega\Delta z\right)}\exp\left(\frac{ia}{2\sin\left(\omega\Delta z\right)}\left[\left(r_{1}^{2}+r_{2}^{2}\right)\cos\left(\omega\Delta z\right)-2\vec{r}_{1}\cdot\vec{r}_{2}\right]\right)e^{ik_{min}\left(z_{2}-z_{1}\right)},\label{eq:W-oscillator}\\
\omega^{2} & = & \frac{-2iC\rho_{A}}{\nu\alpha(1-\alpha)},\nonumber \\
a^{2} & = & -iC\rho_{A}\nu\alpha(1-\alpha)/2\nonumber \end{eqnarray}

\section{Gluon shadowing}

\label{sec:GluonShadowing}

It has been known since~\cite{Kancheli:1973vc,Gribov:1984tu} that
in addition to the quark shadowing inside nuclei there is also shadowing
of gluons, which leads to attenuation of the gluon parton distributions.
While nuclear shadowing of quarks is directly measured in DIS, the
shadowing of gluons is poorly known from data~\cite{Arleo:2007js,Armesto:2006ph},
mainly due to the relatively large error bars in the nuclear structure
functions and their weak dependence on the gluon distributions, which
only comes via evolution. The theoretical predictions for gluon shadowing
strongly depend on the implemented model--while for $x_{B}\gtrsim10^{-2}$
they all predict that the gluon shadowing is small or absent, for
$x_{B}\lesssim10^{-2}$ the predictions vary in a wide range (see
the review~\cite{Armesto:2006ph} and references therein). Some of
the recent analysis \cite{eps08} led to such a strong gluon shadowing,
that the unitarity bound \cite{bound} was severely broken. Since
in this paper we also make predictions for the LHC energy range, the
gluon shadowing corrections should be taken into account as well.

The attenuation factor $R_{g}$ \[
R_{g}\left(x,Q^{2},b\right)=\frac{G_{A}\left(x,Q^{2},b\right)}{T_{A}(b)G_{N}\left(x,Q^{2},b\right)},\]
 where $G_{N}\left(x,Q^{2},b\right)$ is the impact parameter dependent
gluon PDF, was evaluated in the dipole approach in~\cite{kst2,Kopeliovich:2001hf}.
It was found convenient to evaluate $R_{g}$ relating it to the shadowing
corrections in DIS with longitudinally polarized photons, \begin{equation}
R_{g}\left(x,Q^{2},b\right)\approx1-\frac{\Delta\sigma_{L}^{\gamma^{*}p}\left(x,Q^{2},b\right)}{T_{A}(b)\sigma_{L}^{\gamma^{*}p}\left(x,Q^{2}\right)},\label{eq:R_g_damping}\end{equation}
 where $\Delta\sigma_{L}^{\gamma^{*}p}=\sigma_{L}^{\gamma^{*}A}-A\sigma_{L}^{\gamma^{*}p}$
is the shadowing correction at impact parameter $b$, and $\sigma_{L}^{\gamma^{*}p}\left(x,Q^{2}\right)$
is the total photoabsorption~cross section for a longitudinal photon.
The process with longitudinal photons is chosen because the aligned
jets configurations are suppressed by powers of $Q^{2}$, so that
the average size of the dipole is small, $\left\langle r^{2}\right\rangle \sim1/Q^{2},$
and nuclear shadowing mainly originates from gluons.

As it was shown in~\cite{kst2,Kopeliovich:2001hf},

\begin{equation}
\Delta\sigma_{L}^{\gamma^{*}p}\left(x,Q^{2},b\right)=\int_{-\infty}^{+\infty}dz_{1}\int_{-\infty}^{+\infty}dz_{2}\Theta\left(z_{2}-z_{1}\right)\rho_{A}\left(b,z_{1}\right)\rho_{A}\left(b,z_{2}\right)\Gamma\left(x,Q^{2},z_{2}-z_{1}\right),\label{eq:DeltaSigmaGluon}\end{equation}
 where $\rho_{A}(b,z)$ is the nuclear density, and~$\Gamma\left(x,Q^{2},\Delta z\right)$
is defined as \begin{eqnarray*}
\Gamma\left(x,Q^{2},\Delta z\right) & = & \Re e\int_{x}^{0.1}\frac{d\alpha_{G}}{\alpha_{G}}\frac{16\alpha_{em}\left(\sum Z_{q}^{2}\right)\alpha_{s}\left(Q^{2}\right)C_{eff}^{2}}{3\pi^{2}Q^{2}\tilde{b}^{2}}\times\\
 & \times & \left[\left(1-2\zeta-\zeta^{2}\right)e^{-\zeta}+\zeta^{2}(3+\zeta)E_{1}(\zeta)\right]\\
 & \times & \left[\frac{t}{w}+\frac{\sinh\left(\Omega\Delta z\right)}{t}\ln\left(1-\frac{t^{2}}{u^{2}}\right)+\frac{2t^{3}}{uw^{2}}+\frac{t\sinh\left(\Omega\Delta z\right)}{w^{2}}+\frac{4t^{3}}{w^{3}}\right],\end{eqnarray*}
 with \begin{eqnarray*}
\tilde{b}^{2} & = & \left(0.65GeV\right)^{2}+\alpha_{G}Q^{2},\\
\Omega & = & \frac{iB}{\alpha_{G}\left(1-\alpha_{G}\right)\nu},\\
B & = & \sqrt{\tilde{b}^{4}-i\alpha_{G}\left(1-\alpha_{G}\right)\nu C_{eff}\rho_{A}},\\
\nu & = & \frac{Q^{2}}{2m_{N}x},\\
\zeta & = & ixm_{N}\Delta z,\\
t & = & \frac{B}{\tilde{b}^{2}},\\
u & = & t\cosh\left(\Omega\Delta z\right)+\sinh\left(\Omega\Delta z\right),\\
w & = & \left(1+t^{2}\right)\sinh\left(\Omega\Delta z\right)+2t\cosh\left(\Omega\Delta z\right).\end{eqnarray*}
Notice the importance of the $Q$-independent term in $\tilde{b}^{2}$,
which controls the mean quark-gluon, or glue-glue separation at low
scale. The magnitude of this term dictated by various experimental
data \cite{spots} and especially by data on diffraction \cite{kst2},
is rather large $b_{0}^{2}=(0.65\, GeV)^{2}$. This leads to a small
dipole sizes $r_{0}\sim0.3$fm and weak gluon shadowing.

For heavy nuclei we may rely on the hard sphere approximation, $\rho_{A}(r)\approx\rho_{A}(0)\Theta\left(R_{A}-r\right)$,
and simplify~(\ref{eq:DeltaSigmaGluon}) to: \begin{eqnarray*}
\Delta\sigma^{\gamma^{*}p}\left(x,Q^{2},b\right) & \approx & \rho_{A}^{2}(0)\int_{0}^{L}d\Delta z(L-\Delta z)\Gamma\left(x,Q^{2},\Delta z\right),\end{eqnarray*}
 where $L=2\sqrt{R_{A}^{2}-b^{2}}.$ For the total cross-section after
integration over $\int d^{2}b$ we get \begin{eqnarray*}
\Delta\sigma^{\gamma^{*}p}\left(x,Q^{2}\right) & = & \int d^{2}b\Delta\sigma^{\gamma^{*}p}\left(x,Q^{2},b\right)\\
 & \approx & \frac{\pi\rho_{A}^{2}(0)}{12}\int_{0}^{2R}d\Delta z\Gamma\left(x,Q^{2},\Delta z\right)\left(16R_{A}^{3}-12R_{A}^{2}\Delta z+\Delta z^{3}\right)\end{eqnarray*}
 The results of the evaluation of the gluon shadowing are presented
in Section~\ref{sec:Results}.

\section{Wave functions from the instanton vacuum}

\label{sec:WFfromIVM}In this section we present briefly some details
of the evaluation of the photon wavefunction in the instanton vacuum
model~(see~\cite{Schafer:1996wv,Diakonov:1985eg,Diakonov:1995qy}
and references therein). The central object of the model is the effective
action for the light quarks in the instanton vacuum, which in the
leading order in $N_{c}$ has the form~\cite{Diakonov:1985eg,Diakonov:1995qy}
\[
S=\int d^{4}x\left(\frac{N}{V}\ln\lambda+2\Phi^{2}(x)-\bar{\psi}\left(\hat{p}+\hat{v}-m-c\bar{L}f\otimes\Phi\cdot\Gamma_{m}\otimes fL\right)\psi\right),\]
 where $\Gamma_{m}$ is one of the matrices, $\Gamma_{m}=1,i\vec{\tau},\gamma_{5},i\vec{\tau}\gamma_{5}$,~$\psi$
and $\Phi$ are the fields of constituent quarks and mesons respectively,
$N/V$ is the density of the instanton gas, $\hat{v}\equiv v_{\mu}\gamma^{\mu}$
is the external vector current corresponding to the photon, $L$ is
the gauge factor, \begin{equation}
L\left(x,z\right)=P\exp\left(i\int\limits _{z}^{x}d\zeta^{\mu}v_{\mu}(\zeta)\right),\label{eq:L-factor}\end{equation}
 which provides the gauge covariance of the action, and $f(p)$ is
the Fourier transform of the zero-mode profile.

In the leading order in $N_{c}$, we have the same Feynman rules as
in perturbative theory, but with a momentum-dependent quark mass $\mu(p)$
in the quark propagator\begin{eqnarray}
S(p) & = & \frac{1}{\hat{p}-\mu(p)+i0}.\end{eqnarray}
 The mass of the constituent quark has a form \[
\mu(p)=m+M\, f^{2}(p),\]
 where $m\approx5$~MeV is the current quark mass, $M\approx350$~MeV
is the dynamical mass generated by the interaction with the instanton
vacuum background. Due to the presence of the instantons the coupling
of a vector current to a quark is also modified,\begin{eqnarray}
\hat{v} & \equiv & v_{\mu}\gamma^{\mu}\rightarrow\hat{V}=\hat{v}+\hat{V}^{nonl},\nonumber \\
\hat{V}^{nonl} & \approx & -2Mf(p)\frac{df(p)}{dp_{\mu}}v_{\mu}(q)+\mathcal{O}\left(q^{2}\right).\label{eq:V-nonl-expanded}\end{eqnarray}
 Notice that for an arbitrary photon momentum $q$ the expression
for $\hat{V}^{nonl}$ depends on the choice of the path in~(\ref{eq:L-factor})
and as a result one can find in the literature different expressions
used for evaluations~\cite{Dorokhov:2006qm,Anikin:2000rq,Dorokhov:2003kf,Goeke:2007j}.
In the limit $p\to\infty$ the function $f(p)$ falls off as $\sim\frac{1}{p^{3}},$
so for large $p\gg\rho^{-1}$, where $\rho\approx(600\, MeV)^{-1}$
is the mean instanton size, the mass of the quark $\mu(p)\approx m$
and the vector current interaction vertex $\hat{V}\approx\hat{v}$.
However, we would like to emphasize that the wavefunction $\Psi(\beta,r)$
gets contributions from both the soft and the hard parts, so even
in the large-$Q$ limit the instanton vacuum function is different
from the well-known perturbative result.

We have to evaluate the wavefunctions associated with the following
matrix elements:

\begin{eqnarray}
I_{\Gamma}(\beta,\vec{r}) & = & \int\frac{dz^{-}}{2\pi}e^{i\left(\beta+\frac{1}{2}\right)q^{-}z^{+}}\left\langle 0\left|\bar{\psi}\left(-\frac{z}{2}n-\frac{\vec{r}}{2}\right)\Gamma\psi\left(\frac{z}{2}n+\frac{\vec{r}}{2}\right)\right|\gamma(q)\right\rangle ,\end{eqnarray}
 where $\Gamma$ is one of the matrices $\Gamma=\gamma_{\mu},\gamma_{\mu}\gamma_{5},\sigma_{\mu\nu}.$
In the leading order in $N_{c}$ one can easily obtain\begin{equation}
I_{\Gamma}=\int\frac{d^{4}p}{(2\pi)^{4}}e^{i\vec{p}_{\perp}\vec{r}_{\perp}}\delta\left(p^{+}-\left(\beta+\frac{1}{2}\right)q^{+}\right)Tr\left(S(p)\hat{V}S(p+q)\Gamma\right).\label{eq:WF-LO}\end{equation}
 The evaluation of~(\ref{eq:WF-LO}) is quite tedious but straightforward.
Details of this evaluation may be found in~\cite{Dorokhov:2006qm}.

In what follows, we use Eqn~(\ref{eq:WF-LO}) for the evaluation
of the initial and final wavefunctions $\Psi_{in,f}$. Without any
loss of generality, we may choose the frame with $q=(M_{\bar{l}l},0,0,0)$
for the time-like photon, and the frame where $q=(0,0,0,Q)$ for the
space-like photon, and after that make a boost to the Bjorken frame.
However, we would like to emphasize that application of (\ref{eq:WF-LO})
to the case of time-like photon should be done with care. Since the
instanton vacuum model does not possess confinement, rigorously speaking
Eq.~(\ref{eq:WF-LO}) may be applied only if the virtuality is below
the quark pair production threshold, $M_{\bar{l}l}^{2}\sim4\mu^{2}(0)\sim m_{\rho}^{2}$
(in case of pQCD wave functions the corresponding threshold is located
at $M_{\bar{l}l}^{2}\sim4m_{q}^{2}$). Such behavior is characteristic
to all models which have quarks as degrees of freedom but do not have
built-in confinement.

The overlap of the initial and final photon wavefunctions in~(\ref{eq:DVCS-cross-section-2})
was evaluated according to\begin{equation}
\Psi_{f}^{(i)*}\left(\beta,r,-M_{\bar{l}l}^{2}\right)\Psi^{(i)}\left(\beta,r,Q^{2}\right)=\sum_{\Gamma}I_{\Gamma}^{*}\left(\beta,r^{*},-M_{\bar{l}l}^{2}\right)I_{\Gamma}\left(\beta,r,Q^{2}\right),\label{eq:ConvolutionWF}\end{equation}
 where the summation is over the possible polarization states $\Gamma=\gamma_{\mu},\gamma_{\mu}\gamma_{5},\sigma_{\mu\nu}$.
In the final state we should use $r_{\mu}^{*}=r_{\mu}+n_{\mu}\frac{q_{\perp}'\cdot r_{\perp}}{q_{+}}=r_{\mu}-n_{\mu}\frac{\Delta_{\perp}\cdot r_{\perp}}{q_{+}}$,
which is related to the reference frame with $q'_{\perp}=0$, in which
the components~(\ref{eq:WF-LO}) were evaluated.

\section{Numerical results}

\label{sec:Results}

\begin{figure}
\includegraphics[scale=0.4]{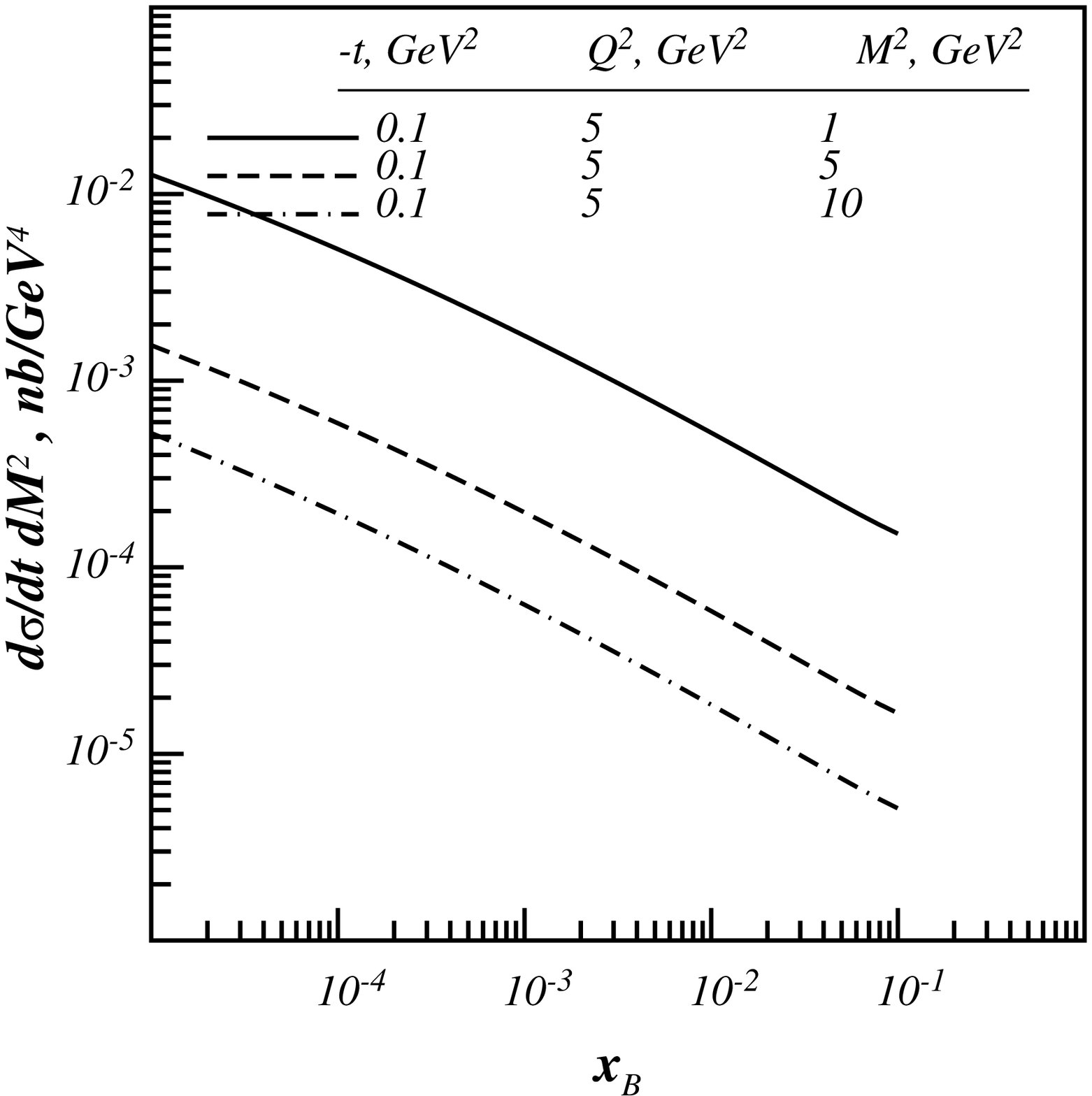}\includegraphics[scale=0.4]{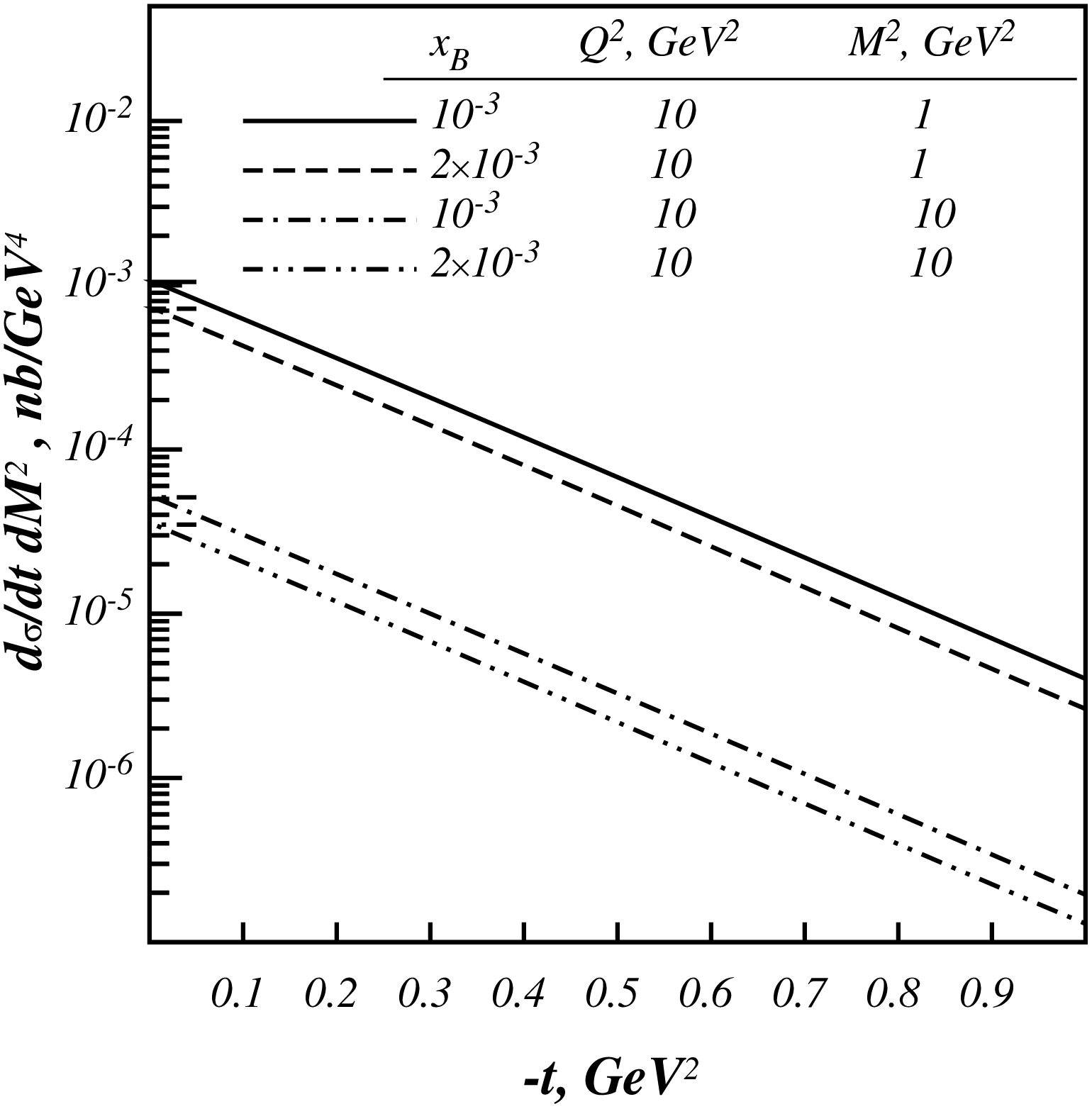}

\caption{\label{fig:DDVCS-proton}DDVCS cross-section on the proton. Left:
$x_{B}$-dependence for different kinematical points. Right: $t$-dependence
for different kinematical points.}

\end{figure}

In this section we present the results of numerical calculations.
While currently there is no data for DDVCS, we expect that similar
to DVCS the experiments will be done in the region of large virtualities
$Q^{2}.$ In that region we have Bjorken scaling, so all the model
parameters such as basic cross-section $\sigma_{0}$ and saturation
radius $R_{0}$ in Eq.~(\ref{eq:GPDParametrization}). should depend
on the Bjorken $x_{B}.$ A widely accepted parameterization which
incorporates this feature is the GBW-type parameterization~\cite{GolecBiernat:1998js,Kopeliovich:2008da,Kopeliovich:2007fv,Kopeliovich:2008nx,Kopeliovich:2009cx}. The DDVCS cross-sections on the proton are shown in the Figure~\ref{fig:DDVCS-proton}.
 Similar to DVCS, the DDVCS cross-section increases in the small-$x$
region like $x^{\alpha}$. As a function of $t$, the cross-section
is exponentially decreasing.

\begin{figure}
\includegraphics[scale=0.4]{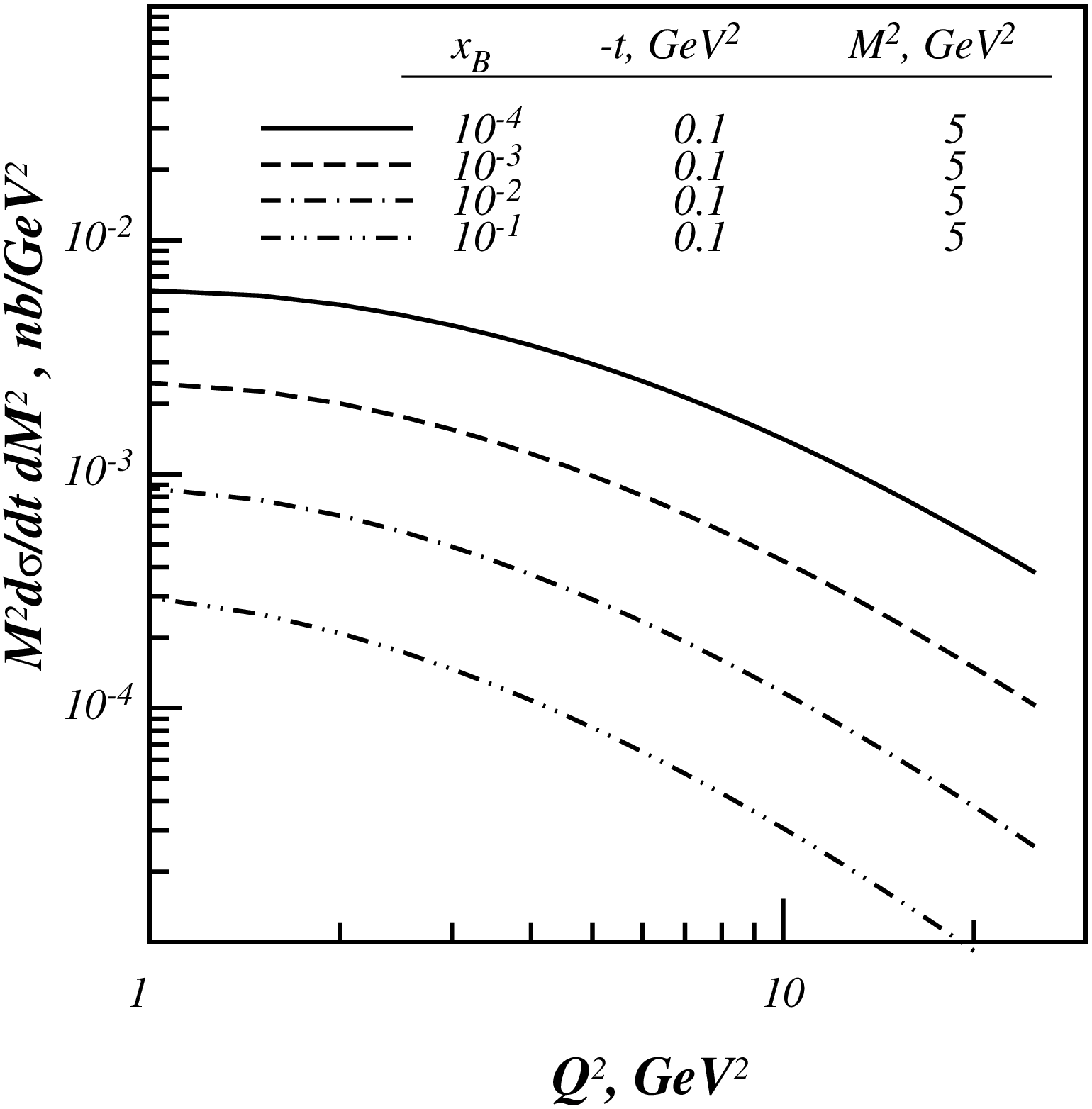}\includegraphics[scale=0.4]{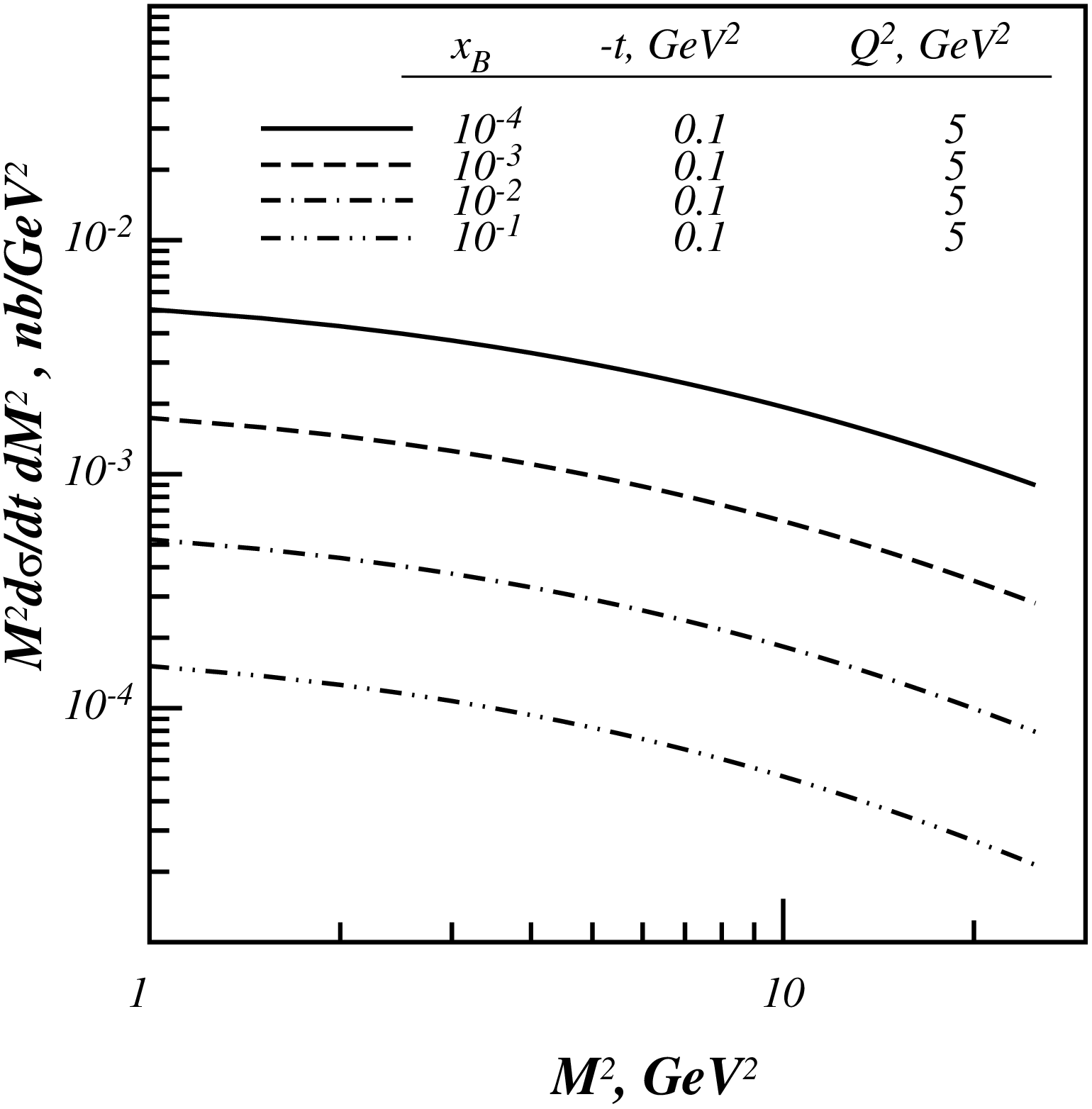}

\caption{\label{fig:DDVCS-Q2M2}DDVCS cross-section on the proton. Here we
plot $M^{2}d\sigma/dt\, dM^{2}$ instead of $d\sigma/dt\, dM^{2}$~in
order to hide the trivial $1/M^{2}$ in~Eq.~(\ref{eq:DDVCS-corss-section}).
Left: $Q^{2}$-dependence for different kinematical points. Right:
$M^{2}$-dependence for different kinematical points.}

\end{figure}

The $Q^{2}$- and $M^{2}$-dependence of the DDVCS cross-section on
the proton is shown in the Figure~\ref{fig:DDVCS-Q2M2}. As a function
of the virtuality $Q^{2}$, the DDVCS cross-section weakly depends
on $Q^{2}$ for $Q^{2}\lesssim M_{\bar{l}l}^{2}$, but decreases like
$Q^{-4}$ for $Q^{2}\gg M_{\bar{l}l}^{2}$. Similar behavior is observed
for $M^{2}d\sigma/dt\, dM^{2}$ for fixed $Q^{2}$. Physically, such
behavior is clear: the average dipole size $\left\langle r^{2}\right\rangle $
is controlled by the wave function with largest virtuality, being
$\left\langle r^{2}\right\rangle \sim1/\max\left(Q^{2},\, M_{\bar{l}l}^{2}\right)$.\\

\begin{figure}
\includegraphics[scale=0.4]{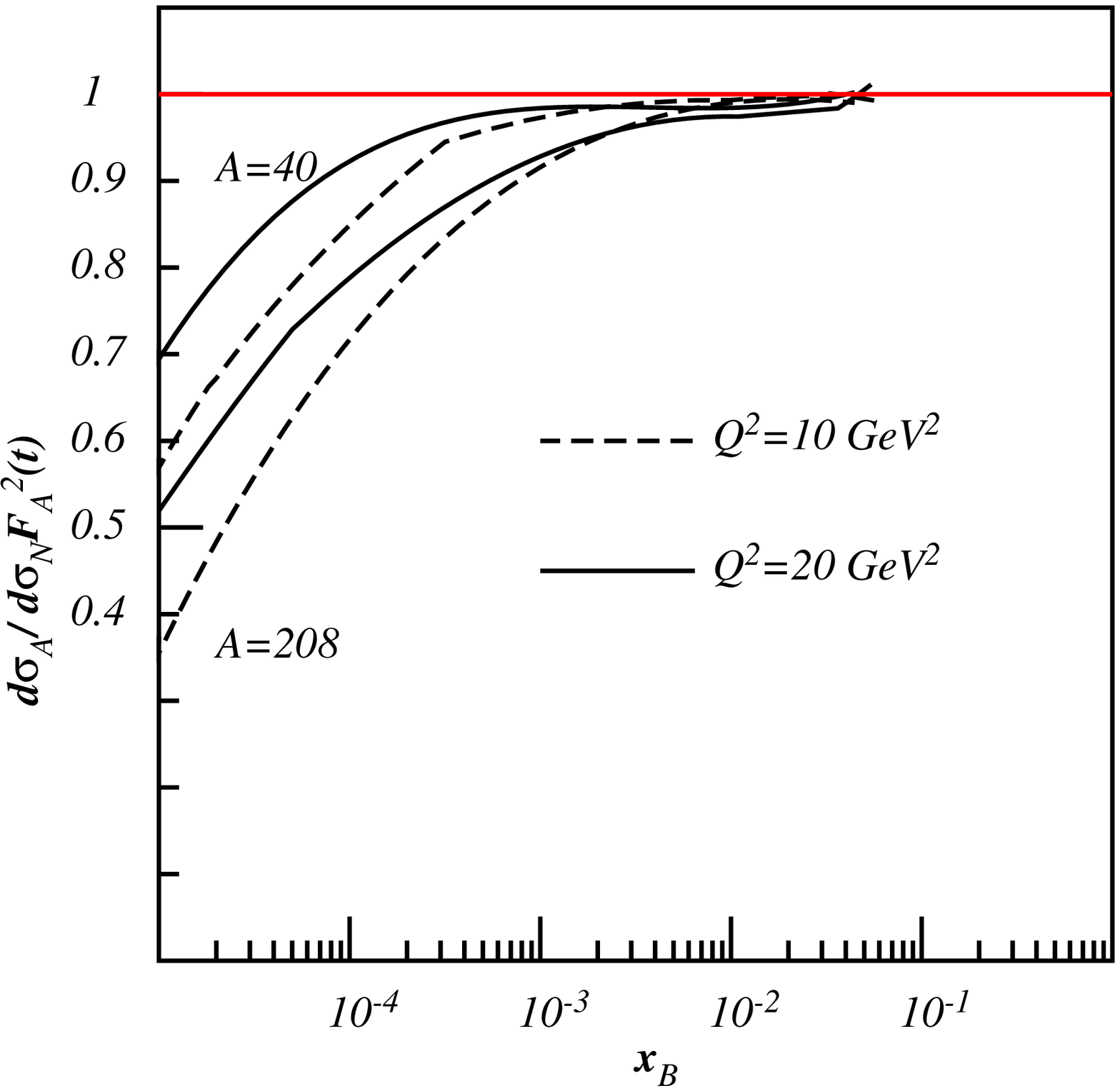}\includegraphics[scale=0.4]{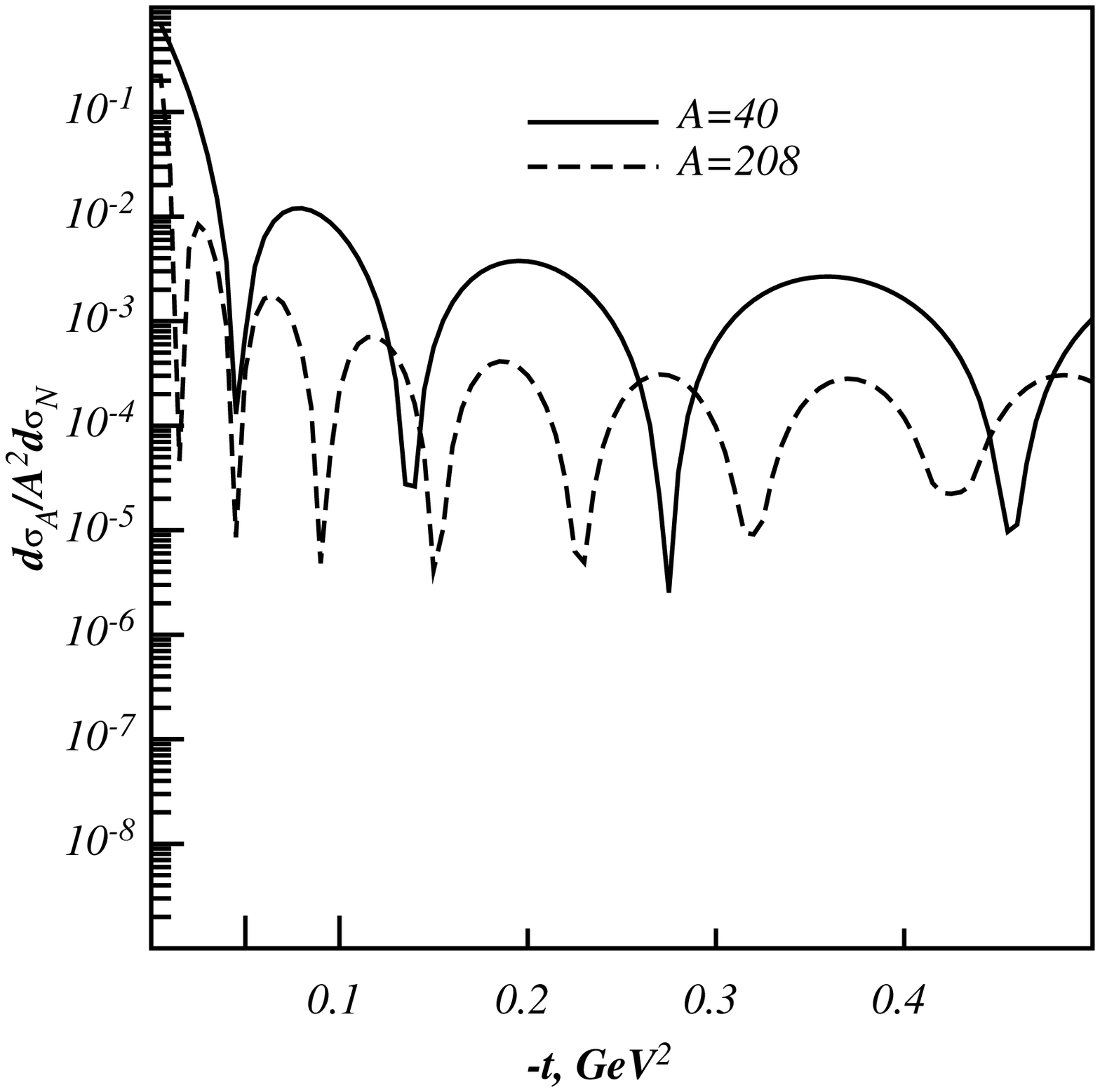}

\caption{\label{fig:DVCS-COH-RES} The nucleus to nucleon cross section ratio
for the DDVCS as function of different kinematical variables. Left:
$x_{B}$-dependence of the shadowing, $t=t_{min}$, $M_{\bar{l}l}^{2}=10$~GeV$^{2}.$
for different $Q^{2}$ and $A$. From top to bottom: $A=40$ and $A=208$.
Right: $t$-dependence, $x_{B}=10^{-3}$, $M_{\bar{l}l}^{2}=10$GeV$^{2}$.}

\end{figure}

As one can see from the Figure~\ref{fig:DVCS-COH-RES}, the shadowing
correction is increasing towards small $x_{B}$, and for $x_{B}\sim10^{-5}$
the nuclear cross-section ratio decreases by a factor of two compared
to the naive estimate $d\sigma_{A}\sim F_{A}^{2}(t)d\sigma_{N}$.
As a function of the momentum transfer $t$, the shadowing correction
reveals a behavior qualitatively similar to the nuclear formfactor
$F_{A}(t)$: it steeply drops at small-$t$ and has zeros for some
$t$. Notice, however, that the zero positions in the cross-section
do not coincide with the zeros of the formfactor. This is a result
of shadowing which suppresses the contribution of the central part
of the nucleus and modifies the $b$-dependence of the cross section
compared to the formfactor.

\begin{figure}
\includegraphics[scale=0.4]{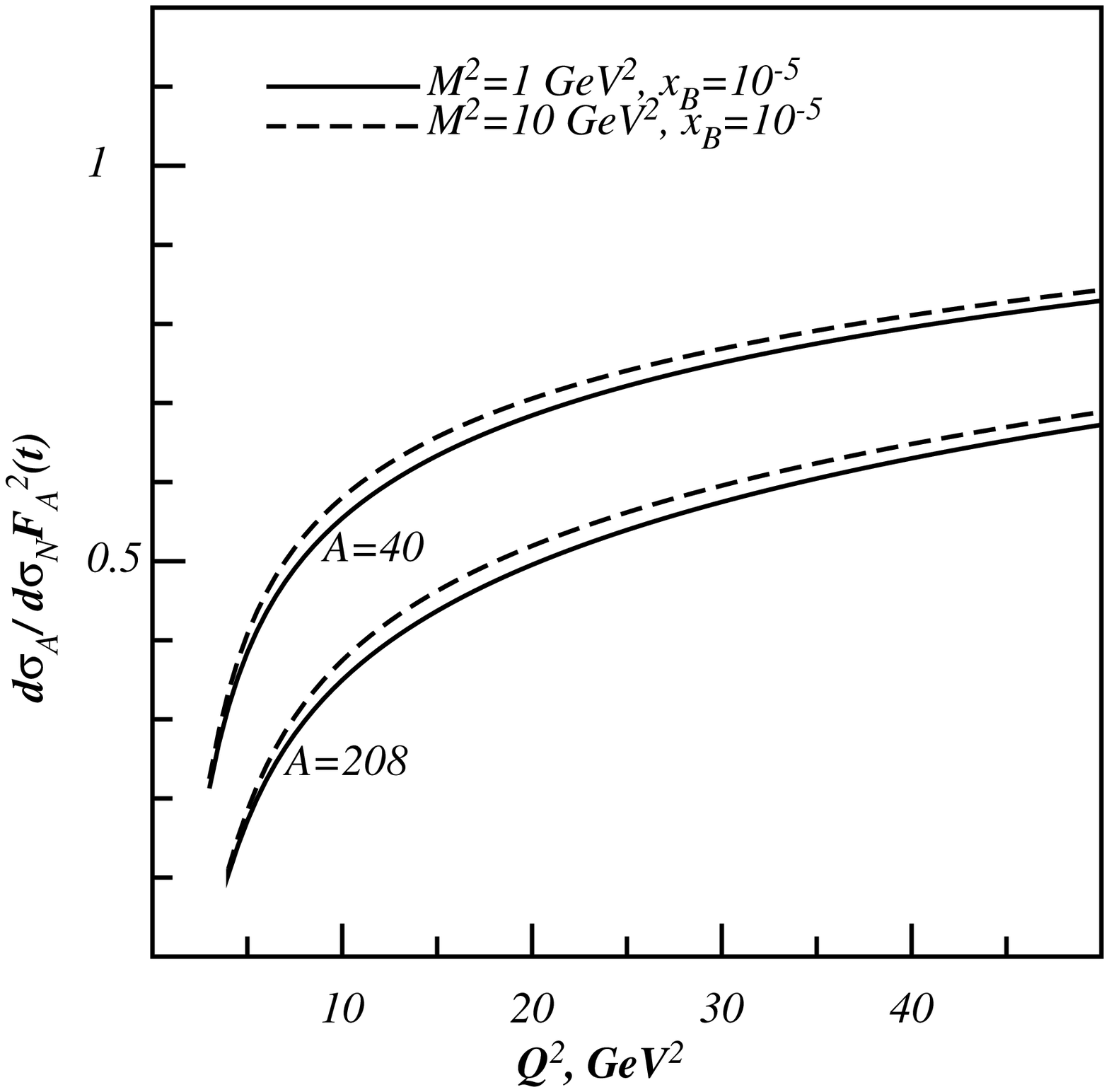}\includegraphics[scale=0.4]{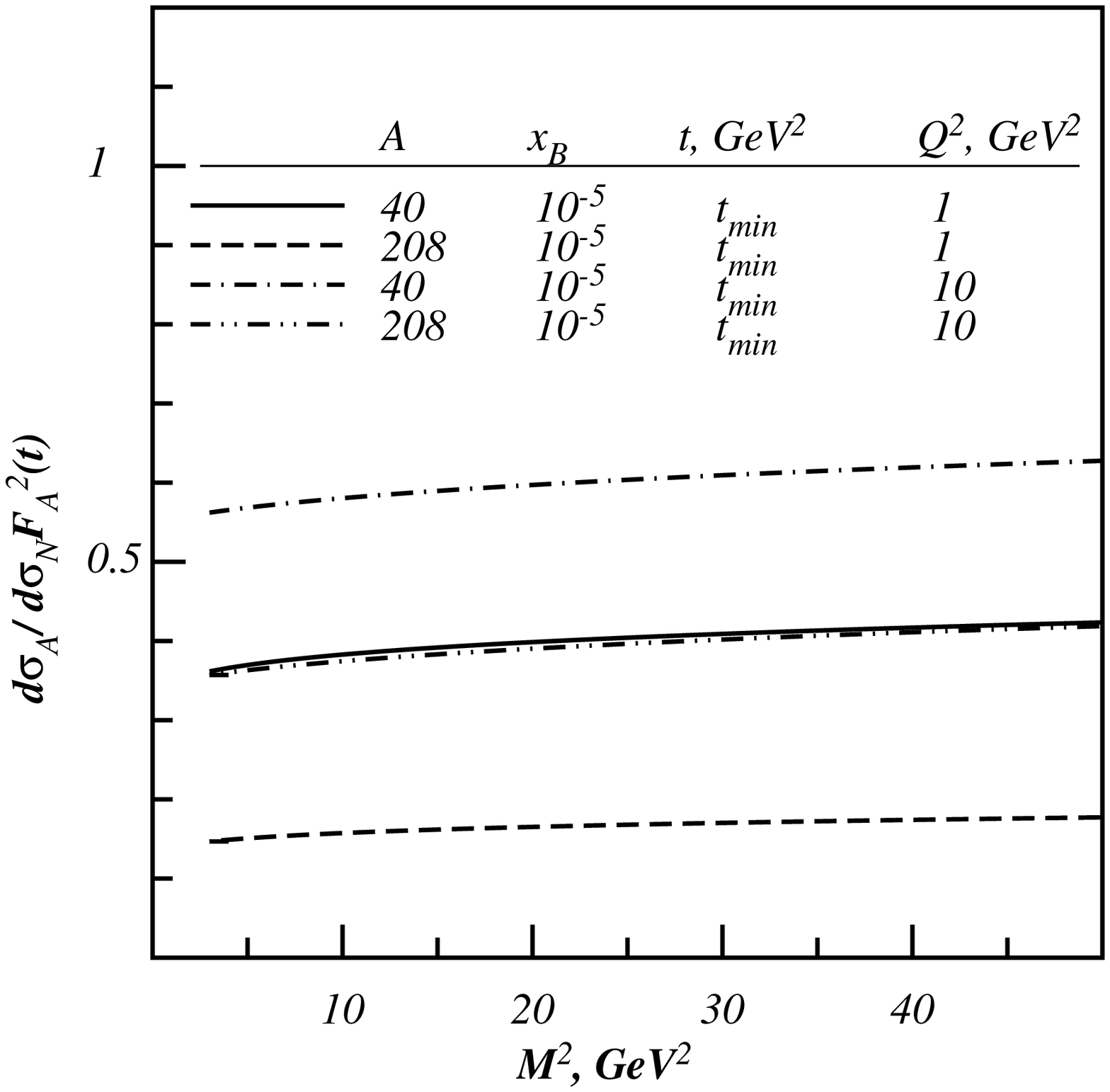}

\caption{\label{fig:DDVCS_NUCL_Q2M2}The nucleus to nucleon cross section ratio
for the DDVCS as function of different kinematical variables. Left:
$Q^{2}$-dependence, $t=t_{min}$,$x_{B}=10^{-5}$ for different $M_{\bar{l}l}^{2}$
and $A$. From top to bottom: $A=40$ and $A=208$. Right: $M^{2}$-dependence
for different kinematical variables.}

\end{figure}

The $Q^{2}$- and $M^{2}$-dependence of the shadowing is shown in
the Figure~\ref{fig:DDVCS_NUCL_Q2M2}. The $Q^{2}$-dependence is
similar to the DVCS case: the shadowing ratio is homogeneously increasing
and for asymptotically large $Q^{2}$ reaches 1. The shadowing is
also increasing as a function of $M^{2}$, but not so fast as a $Q^{2}$-dependence.

Concluding, we considered DDVCS on the proton and nuclear targets
within the color dipole model. We found that the magnitude of the
cross-section is small, of order a few picobarns, and thus requires
accelerators with high luminosities (e.g. future electron-ion colliders
EIC and LHeC,~\cite{Klein:2008zza,Zimmermann:2008zzd}). The nuclear
shadowing in the process is large and is important for analysis of
DDVCS on nuclei.

\section*{Acknowledgments}

This work was supported in part by Fondecyt (Chile) grants 1090291,
1090073, 1100287, and by DFG (Germany) grant PI182/3-1.

\end{document}